\documentclass[aps,prc,superscriptaddress,amsfonts,amssymb,amsmath,bibnotes,footinbib,twocolumn,fleqn,10pt]{revtex4-1}

\usepackage{graphicx}
\usepackage{subfigure}
\usepackage{leftidx}
\usepackage{setspace}
\usepackage{array}
\usepackage{lineno,hyperref}
\usepackage{amssymb}
\usepackage{amsmath}

\begin{document}
\begin{spacing}{1.0}

\title{The Fractal Geometrical Properties of Nuclei}


\author{W.~H.~Ma}
\email[]{weihuma@impcas.ac.cn}
\affiliation{Institute of Modern Physics, Chinese Academy of Science, Lanzhou 730000, People's Reublic of China}
\affiliation{University of Chinese Academy of Science, Beijing, 100049, People's Reublic of China}
\author{J.~S.~Wang}
\affiliation{Institute of Modern Physics, Chinese Academy of Science, Lanzhou 730000, People's Reublic of China}
\author{Q.~Wang}
\affiliation{Institute of Modern Physics, Chinese Academy of Science, Lanzhou 730000, People's Reublic of China}
\author{S.~Mukherjee}
\affiliation{Physics Department, Faculty of Science, M.S. University of Baroda, Vadodara - 390002, India}
\author{L.~Yang}
\affiliation{Institute of Modern Physics, Chinese Academy of Science, Lanzhou 730000, People's Reublic of China}
\author{Y.~Y.~Yang}
\affiliation{Institute of Modern Physics, Chinese Academy of Science, Lanzhou 730000, People's Reublic of China}
\author{M.~R.~Huang}
\affiliation{Institute of Modern Physics, Chinese Academy of Science, Lanzhou 730000, People's Reublic of China}
\author{Y.~J.~Zhou}
\affiliation{Institute of Modern Physics, Chinese Academy of Science, Lanzhou 730000, People's Reublic of China}
\affiliation{University of Chinese Academy of Science, Beijing, 100049, People's Reublic of China}


\date{\today}

\begin{abstract}
We present a new idea to understand the structure of nuclei, which is comparing to the liquid drop model. After discussing the probability that the nuclear system may be a fractal object with the characteristic of self-similarity, the nuclear irregular structure properties and the self-similarity characteristic are considered to be an intrinsic aspects of nuclear structure properties. For the description of nuclear geometric properties, nuclear fractal dimension is an irreplaceable variable similar to the nuclear radius. In order to determine these two variables, a new nuclear potential energy formula which is related to the fractal dimension is put forward and the phenomenological semi-empirical Bethe-Weizs$\ddot{a}$cker binding energy formula is modified using the fractal geometric theory. And one important equation set with two equations is obtained, which is related to the conception that the fractal dimension should be a dynamical parameter in the process of nuclear synthesis. The fractal dimensions of the light nuclei are calculated and their physical meanings are discussed. We have compared the nuclear fractal mean density radii with the radii calculated by the liquid drop model for the light stable and unstable nuclei using rational nuclear fractal structure types. In the present model of fractal nuclear structure there is an obvious feature comparing to the liquid drop model, since the present model can reflect the geometric informations of the nuclear structure, especially for the nuclei with clusters, such as the $\alpha$-cluster nuclei and halo nuclei.
\end{abstract}


\maketitle

\section{Introduction}
Both in theory \cite{bibitem1,bibitem2,bibitem3,bibitem4,bibitem5} and experiment \cite{bibitem6,bibitem7,bibitem8,bibitem9,bibitem10,bibitem11,bibitem12}, the concept of the cluster structure plays an important role in nuclear reaction, nuclear structure and the nuclear excitation. The nuclear distribution is non-uniform especially for halo nuclei. For instance, ${}^{11}Li$ has a core and valence nucleons (${}^{9}Li+n+n$) halo structure and the matter distribution is much non-uniform \cite{bibitem10,bibitem13}. Its radius is much larger than given by the usual expression ($r_{0}A^{\frac{1}{3}}$), which depends on the uniform-density liquid drop model.

In addition, the phenomenological semi-empirical Bethe-Weizs$\ddot{a}$cker binding energy formula \cite{bibitem14,bibitem15,bibitem16} for the masses of nuclei has been derived from the liquid drop model and successfully used to calculate the binding energy for stable nuclei and the nuclei very near the stable line. However, there are two points we cannot ignore. On one hand, the liquid drop model for calculating nuclear binding energy strictly depends on the experimental data. On the other hand, it is difficult to describe the light and halo nuclei because it depends on statistic and cannot reflect the structure properties in nucleus precisely. Taking into consideration the fact that, due to the nuclear particle property and its quantum motion, it is not proper to regard nucleus as a compact sphere with smooth surface and there is no explicit border. The nucleons in a nucleus are separated and there are void spaces among them. These void spaces near the surface are much bigger in halo or weakly-bounded nuclei in comparison to that in stable nuclei. The notion of radii is only a statistical average effect of nuclear matter distribution. In fact, the nuclear structures are irregular, which are the obvious features different from the description of the liquid drop model.

In order to describe the nuclear structure properties reasonably well, the notion of fractal object for the irregular systems with non-integral dimensions is possible, which was introduced by Mandelbrot in science in 1967 \cite{bibitem17}. The self-similarity and the scale invariance are the most important characteristics of a fractal object. The fractal self-similarity means that the part of the fractal object is similar to the whole fractal object after magnification. For an arbitrary part of the fractal object, its some important properties, such as shape, complexity and irregularity, keep invariant after magnifying or shrinking certain proportion, which is the meaning of scale invariance. For a regular fractal object, there is no characteristic size but characteristic fractal dimension. However, the self-similarity of an approximate or statistical fractal object exists in a finite scale range. In nuclear system with clusters structure, due to the similar nuclear and electromagnetic interaction between the clusters in a nucleus, it is possible that the geometric structure and the physical laws are similar between the nucleons in cluster and the clusters in nucleus and also between the clusters and the whole nucleus. The nucleus can be deemed as a statistical self-similarity fractal system with finite scale range. Xavier Campi had found the existence of finite size scaling in nuclear fragmentation \cite{bibitem18}. He introduced the fractal dimension which give the information on the internal structure of the fragments during the fragmentation process. Adamenko et al \cite{bibitem19} queried the liquid-like model and studied the properties in super-heavy nuclear isomers using fractal theory. These studies are the fine enlightenment of connecting the characteristics of the nuclear structure with the fractal geometric theory.

As explained above, the nuclear irregular structure properties and the self-similarity characteristic may be an intrinsic aspects of nuclear structure properties. It is possible to use the fractal theory to describe the nuclear structure properties. The objective of the present work is to introduce this new concept of nuclear fractal structures to study the nuclear properties.

\section{Description of the model of nuclear fractal structure}
In our present description, a more general conception of nuclear fractal clusters (NFCs) is applied, which is similar but different from the conventional one of the $\alpha$-cluster structure and the core plus valence nucleons structure in halo nuclei. The latter is considered to be one kind of the former which is basing on the concept of the characteristics of fractal objects. The concept of NFCs is that of the nucleus as a fractal assembly of structural subunits that are themselves made up of no less than one nucleon and keep certain correlation of the similarity with the ensemble in the geometrical and in physics. And the geometrical boundary of the NFCs within some nuclei is less distinguishable than that of the clusters in $\alpha$-cluster nuclei and the halo nuclei.

Similar to the definition given in \cite{bibitem20}, an isotropic self-similarity nuclear fractal dimension $D_{f}$ is defined in the following relation \begin{equation} M(b\cdot{r}) = b^{D_{f}}\cdot{M(r)} \end{equation}
where $M(r)$ is the mass number within the size $r$ of the fractal object; $M(b\cdot{r})$ is the mass number of $b$ times the size $r$ of the fractal object, where $b$ is a scaling factor among the similar parts within the fractal object. The only solution for relation (1) is $M(r)\propto{r^{D_{f}}}$. The nuclear average matter density $\rho(r)$ with the law of decay of isotropic spatial correlation, $\rho(r)=\frac{M(r)}{V(r)}\varpropto{\frac{r^{D_{f}}}{r^{3}}}\varpropto{r^{D_{f}-3}}$ , is a basic variable function in nucleus. So far, the geometric dimension of nuclei is considered as 3, because of the concept of liquid drop model. For a real physical nuclear object embedded in 3-dimension Euclidean-space, its dimension must be less than or equal to 3. Most of the nuclei with fractal dimensions approaching 3 are stable, which are more like liquid drops.

In relation (2), we assume that there are several NFCs and $A_{i}$ is the mass number of a NFC within a nucleus with mass number $A$. $\rho_{dis}(r)$ is the density distribution function. $\rho_{i}$ is the fractal mean density of a NFC and $R_{i}$  is the fractal mean density radius of it. $F$ is the number of the NFCs in a given nucleus.
\begin{eqnarray} \rho & = & \frac{A}{V}=\frac{\int{\rho_{dis}(r)\cdot{dv}}}{\int{dv}}=\frac{\sum_{i=1}^{F}A_{i}}{V} \nonumber \\ & = & \frac{\sum_{i=1}^{F}\int\rho_{dis}(r_{i})\cdot{dv_{i}}}{V}
 = \frac{\sum_{i=1}^{F}\rho_{i}\cdot{V_{i}}}{V}
\nonumber \\ & = & \sum_{i=1}^{F}\frac{V_{i}}{V}\cdot{\rho_{i}}=\sum_{i=1}^{F}(\frac{R_{i}}{R})^{3}\cdot{\rho_{i}} \end{eqnarray}
where, $R$ is the nuclear fractal mean density radius defined by the relation $A=\frac{4}{3}\pi{R^{3}}\rho$. And $\rho_{i}=\frac{3A_{i}}{4\pi{R_{i}^{3}}}$.

In addition, due to the assumption of nuclear fractal structure and the mass-radius relation of self-similarity fractal object, we list several basic relations:\begin{equation} A=\sum_{i=1}^{F}A_{i}; \end{equation}\begin{equation} A\propto{R^{D_{f}}};A_{i}\propto{R_{i}^{D_{f}}}. \end{equation}Using relations (3) and (4), we get the relation among $R$, $R_{i}$ and $D_{f}$:\begin{equation} R^{D_{f}}=\sum_{i=1}^{F}R_{i}^{D_{f}}. \end{equation} Due to the two proportional relations in (4), the relation among $R$, $R_{i}$, $A$, $A_{i}$ and $D_{f}$ is \begin{equation} R=R_{i}(\frac{A}{A_{i}})^{\frac{1}{D_{f}}}. \end{equation}Substitute (6) into (2), then get \begin{equation} \rho=\sum_{i=1}^{F}(\frac{A_{i}}{A})^{\frac{3}{D_{f}}}\rho_{i}. \end{equation}And the relation between the NFCs in nucleus is\begin{equation} \rho_{i}=\rho_{j}(\frac{A_{i}}{A_{j}})^{\frac{D_{f}-3}{D_{f}}} \end{equation}

Therefore, to describe the nuclear geometric properties, the nuclear fractal dimension is an irreplaceable variable similar to the nuclear radius. For given nuclear system with cluster structure, the final free variables are ($R$, $D_{f}$) or ($\rho$, $D_{f}$ ), which are also ($R_{i}$, $D_{f}$) or ($\rho_{i}$, $D_{f}$) due to the relations (6) and (7). In order to determine these two variables, we put forward a new nuclear potential energy formula which is related to fractal dimension. Then the phenomenological semi-empirical Bethe-Weizs$\ddot{a}$cker binding energy formula is modified and the total potential energy is obtained. And one important equation set with two equations is obtained, which is related to the conception that the fractal dimension should be a dynamical parameter in the process of nuclear synthesis. So, the calculations of nuclear fractal dimensions and radii can be done.

In nuclear system, considering the self-similarity properties of the nuclear fractal system, we put forward a nuclear potential energy formula\begin{equation} u(r)=\frac{v_{0}D_{f}}{3(D_{f}-2)}(\frac{r}{r_{s}})^{D_{f}-3}, 2<D_{f}\leq{3}. \end{equation} Which is proportional to the nuclear average density $\rho(r)$. $2r_{s}$stands for the minimum scale size of a nuclear fractal system and it is also the maximum size of the minimum cluster element. $v_{0}$ is a coefficient and keeps constant, which is also corresponding to the estimation of the depth of nuclear potential well in liquid drop model when $D_{f}=3$.

From (9) and using the idea of self-similarity, the relation for the nuclear potential energy of one NFC within nucleus is obtained:\begin{equation} u_{i}(r_{i})=\frac{v_{0}D_{f}}{3(D_{f}-2)}(\frac{r_{i}}{r_{s}})^{D_{f}-3}, 2<D_{f}\leq{3} \end{equation}And because of $u(R)\propto{\rho(R)}$ ,$u_{i}(R_{i})\propto{\rho_{i}(R_{i})}$  and (7), the relation between $u(R)$  and $u_{i}(R_{i})$ is\begin{equation} u(R)=\sum_{i=1}^{F}(\frac{A_{i}}{A})^{\frac{3}{D_{f}}}u_{i}(R_{i}). \end{equation}

Next, we modify the phenomenological semi-empirical Bethe-Weizs$\ddot{a}$cker binding energy formula with the fermi gas model and the fractal theory. Here we mainly concern the liquid drop energy and put aside the correction term basing on the microscopic method, such as the description in \cite{bibitem21}. The original one derived from liquid drop model is


\begin{equation} \begin{aligned} B=(u_{depth}-c_{v}-c_{as}(1-\frac{2Z}{A})^{2})A-c_{surf}A^{\frac{2}{3}}\\-c_{Q}\frac{Z(Z-1)}{A^{\frac{1}{3}}}+c_{p}\frac{(-1)^{Z}+(-1)^{A-Z}}{2A^{\frac{4}{3}}};
\end{aligned} \end{equation}where $u_{depth}\approx{58}$ \cite{bibitem22}(the estimation of the depth of nuclear potential well); $c_{v}=42.27$; $c_{as}=23.48$; $c_{surf}=17.72$; $c_{Q}=0.72$; $c_{p}=19.39$. We use the experimental mass data \cite{bibitem23} to fit the other parameters.

The modified formulae are:\begin{equation}  B_{strong}=(v_{depth}-c_{1}(\rho)-c_{2}(\rho)(1-\frac{2Z}{A})^{2})A; \end{equation}
\begin{equation} B_{surf}=-c_{s}4\pi{R^{2}}; \end{equation}
\begin{equation} B_{Q}=-\frac{3}{5}\frac{Z(Z-1)e^{2}}{R}; \end{equation}
\begin{equation} B_{p}=c_{p}\frac{(-1)^{Z}+(-1)^{A-Z}}{2A^{\frac{4}{3}}}; \end{equation}
\begin{equation} B=B_{strong}+B_{surf}+B_{Q}+B_{p}; \end{equation}where, $
c_{1}(\rho)=\frac{3}{5}\varepsilon(\rho)\rho^{\frac{2}{3}}$; $c_{2}(\rho)=\frac{1}{3}\varepsilon(\rho)\rho^{\frac{2}{3}}$; $c_{s}=0.98MeV\cdot{fm^{2}}$; $c_{p}=19.39MeV$; $\varepsilon(\rho)=\varepsilon_{0}(\frac{\rho}{\rho_{0}})^{\alpha}$; $\varepsilon_{0}=264.12MeV^{-1}$; $\rho_{0}=0.138fm^{-3}$. The formula (13) is the volume energy consisting of three parts. The first part is the depth of nuclear potential energy, which is not a constant, but depends on the structure type. The second part and the third part derived from fermi-gas model are the kinetic energy of the nuclear system, because of which the binding energy decrease. $\varepsilon(\rho)$ shares the kinetic energy, which decrease the contribution of kinetic energy. $\varepsilon_{0}$ was explained as a constant related to the virial coefficient in \cite{bibitem19}. Since we consider that the nuclear system has fractal structure that is much different from the fermi gas, $\varepsilon_{0}$ cannot be kept constant any more, but depends on the mean density that is related to the mean density of every cluster in nucleus as relation (7) shows. In the formula of $\varepsilon(\rho)$, $\alpha$ is a parameter. The formulas(14) and (15) for surface energy and the energy because of Coulomb interaction dose not have any parameters. $c_{s}$, the coefficient of surface tension, is about $1MeV/fm^{2}$. Here we still keep the pair energy because the binding energy is larger when the proton and neutron numbers are even and it is smaller when one of the numbers are odd and even more so when both are odd, which maybe has nothing to do with the nuclear fractal structure. Thus all the terms contributing to binding energy are depended on the fractal dimension $D_{f}$ except the pair term. When $D_{f}\rightarrow{3}$ , the modified formula degenerates to the ordinary binding energy formula (12), and then the state of nucleus shifts from the non-uniform density, which have NFCs, to the uniform state of liquid drop model. So, the parameters $c_{s}$, $c_{p}$, $\varepsilon_{0}$ and $\rho_{0}$ can be all derived from (12-17) When $D_{f}\rightarrow{3}$.

A possible formula for the depth of nuclear potential well in the nuclear fractal system is found to be \begin{eqnarray} v_{depth}=\frac{1}{2}(1-(\frac{A-(1+s)Z}{A})^{2})\nonumber\\\times{\sum_{i=1}^{F}\frac{A_{i}}{A}(58+3u_{i}(R_{i})\frac{D_{f}-2}{D_{f}})}  \end{eqnarray}which depends on the nuclear fractal structure. In fact, the depth of nuclear potential well cannot be $58Mev$ for all nuclei because of the different fractal structure for different nuclei. The product factor is introduced to describe the difference between the number of neutrons and protons in nucleus, and $s=\frac{N_{s}}{Z_{s}}$, $N_{s}$ and $Z_{s}$ are the number of neutrons and protons in nucleus on stable line, which makes sure the depth of nuclear potential well get the maximum value for the nuclei on stable line. For the light stable nuclei, $s\approx{1}$. And for the nuclei ${}^{4}He$, ${}^{8}Be$, ${}^{12}C$, ${}^{16}O$, ${}^{20}Ne$, ${}^{24}Mg$, ${}^{28}Si$, ${}^{32}S$, ${}^{36}Ar$ and ${}^{40}Ca$, $s=1$. When $D_{f}\rightarrow{3}$, $u=u_{i}=v_{0}=-u_{depth}.$
\begin{figure}
\centering
\includegraphics[width=0.32\textwidth]{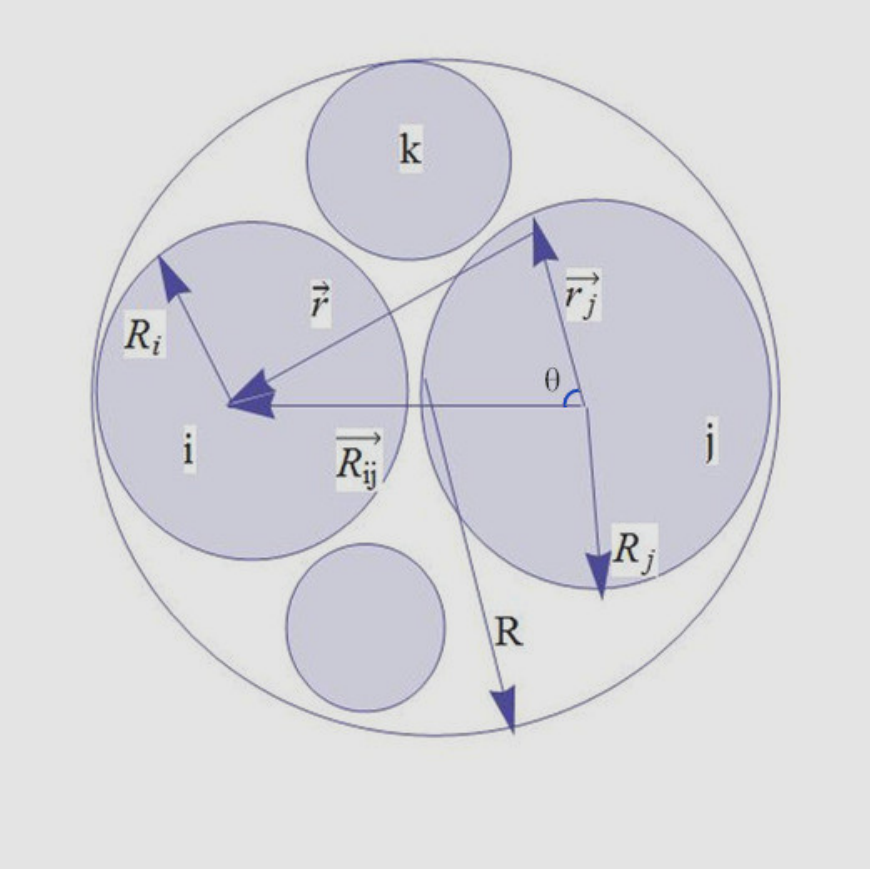}
 \caption{\label{Fig_1} The interaction among clusters in given nucleus. The notations i, j, k stands for different NFCs.}
\end{figure}

We define the total potential energy $U=U(A,Z,D_{f},\rho)$, which is sum of the total nuclear potential energy $U_{A}=U_{A}(A,D_{f},\rho)$ and the total electromagnetic potential energy $U_{Z}=U_{Z}(Z,D_{f},\rho)$. Namely, \begin{equation} U=U_{A}+U_{Z}. \end{equation}The interaction among clusters in given nucleus is showed in FIG. 1. The total nuclear potential energy is \begin{equation} U_{A}=\sum_{i=1}^{F}(\frac{A_{i}}{A})^{\frac{3}{D_{f}}}U_{i}+\sum_{i=1}^{F}\sum_{i\neq{j}}^{F}(\frac{A_{j}}{A})^{\frac{3}{D_{f}}}U_{ij}, \end{equation} Where,

\begin{eqnarray} U_{i}&=& 4\pi\frac{v_{0}D_{f}}{3(D_{f}-2)}\int_{0}^{R_{i}}(\frac{r_{i}}{r_{s}})^{D_{f}-3}\rho_{dis}(r_{i})r_{i}^{2}dr_{i} \nonumber \\
&=& \frac{4\pi{v_{0}\rho_{i}R_{i}^{D_{f}}}}{3(D_{f}-2)r_{s}^{D_{f}-3}}. \end{eqnarray}

\begin{eqnarray} U_{ij}&=& 2{\pi}\int_{0}^{\pi}\int_{0}^{R_{j}}u_{i}(r)\rho_{dis}(r_{j})r_{j}^{2}\sin(\theta)dr_{j}d\theta \nonumber \\
&=& 2\pi\frac{v_{0}D_{f}}{3(D_{f}-2)}\frac{\rho_{j}(R_{ij}+R_{j})^{D_{f}+1}G(D_{f})}{(D_{f}+D_{f}^{3})R_{ij}(R_{ij}^{2}-R_{j}^{2})r_{s}^{D_{f}-3}}.\nonumber \\ \end{eqnarray}
In(22), $ G(D_{f})=(R_{ij}^{2}+D_{f}R_{j}^{2})((\frac{R_{ij}-R_{j}}{R_{ij}+R_{j}})^{D_{f}+1}-1)+(D_{f}+1)R_{ij}R_{j}((\frac{R_{ij}-R_{j}}{R_{ij}+R_{j}})^{D_{f}+1}+1)$; the distance    $r=\sqrt{r_{j}^{2}+R_{ij}^{2}-2r_{j}R_{ij}\cos{\theta}}$ as showed in FIG. 1; $R_{ij}\approx{R_{i}+R_{j}}.$ The total electromagnetic potential energy is \begin{equation} U_{Z}=\sum_{i=1}^{F}U_{Z_{i}}+ \sum_{i=1}^{F}\sum_{i\neq{j}}^{F}U_{Z_{ij}} \end{equation} where,\begin{equation} U_{Z_{i}}=\frac{3}{5}Z_{i}(Z_{i}-1)\frac{e^{2}}{R_{i}}; \end{equation}\begin{equation} U_{Z_{ij}}=\frac{Z_{i}Z_{j}e^{2}}{R_{ij}} \end{equation}

For a given nucleus with NFCs in it, we consider that its fractal dimension $D_{f}$ is constant. However it should be a dynamical parameter in the process of nuclear synthesis. In a given nuclear reaction the binding energy $B=B(A,Z,D_{f},\rho)$ and the total potential energy $U=U(A,Z,D_{f},\rho)$ are changing with $D_{f}$. When $D_{f}$ gets the fixing value, then $B=B(A,Z,D_{f},\rho)$ and $U=U(A,Z,D_{f},\rho)$ get the minimum value, which corresponds to an interacting system becoming relatively stable. So one equation set is gotten: \begin{equation}
\left\{\begin{array}{c}
    \partial_{D_{f}}B(A,Z,D_{f},\rho)=0\\
    \partial_{D_{f}}U(A,Z,D_{f},\rho)=0 .
  \end{array}\right.
 \end{equation}
Because of the uncertain parameter $\alpha$, we need more than two equations to study the structure properties in nucleus. The additional one is \begin{equation} B(A,Z,D_{f},\rho)=B_{exp}+E_{excited}, \end{equation}where $B_{exp}$ is the experimental value of binding energy. And the $E_{excited}$ is the change of the binding energy due to the nucleus being excited from the ground state, which is corresponding to the situation that the NFCs structure is forming in the excited nucleus. If only the nuclei in ground states are considered, $E_{excited}=0$.

Finally, we arrive at the modified binding energy formula $B(A,Z,D_{f},\rho)$ and the total potential energy                   $U(A,Z,D_{f},\rho)$, which are the functions of $A$, $Z$, $D_{f}$ and $\rho$. And the important equation set (26) is gotten. Use (26), (27) and assume rational fractal structure types in nucleus, the nuclear fractal dimension and radius can be determined.
\section{Results and discussions}
In the present work, we have done some calculations and obtained some results for light nuclei in ground states. Combining the analyses given in \cite{bibitem1,bibitem2,bibitem3,bibitem4,bibitem5,bibitem6,bibitem7,bibitem8,bibitem9,bibitem10,bibitem11,bibitem12} and \cite{bibitem24,bibitem25,bibitem26} and the consideration that the rational fractal structure of one nucleus is determined by making the seperated energy of NFCs be minimum possibly, the possible NFC structures for these nuclei are shown in TABLE I. We consider that the scale $r_{s}$ is about 1.25 fm, which is the half of the maximum size of one nucleon. And we compare the radii ($R_{L}=r_{0}A^{\frac{1}{3}}$,$r_{0}=1.25fm$) from liquid drop model with the nuclear fractal mean density radii.

\begin{table}
\centering
\begin{tabular}{cccccccc}
\hline
Nuclei & NFCs & Nuclei & NFCs \\
\hline
${}^{5}He$ & ${}^{4}He$+n & ${}^{14}B$ & ${}^{13}B$+n\\
${}^{6}He$ & ${}^{4}He$+n+n & ${}^{15}B$ & ${}^{14}B$+n\\
${}^{7}He$ & ${}^{6}He$+n & ${}^{16}B$ & ${}^{15}B$+n\\
${}^{8}He$ & ${}^{6}He$+n+n & ${}^{9}C$ & ${}^{8}B$+p\\
${}^{5}Li$ & ${}^{4}He$+p & ${}^{10}C$ & ${}^{8}Be$+p+p\\
${}^{6}Li$ & ${}^{4}He$+${}^{2}H$ & ${}^{11}C$ & ${}^{7}Be$+${}^{4}He$\\
${}^{7}Li$ & ${}^{6}Li$+n & ${}^{12}C$ & ${}^{8}Be$+${}^{4}He$\\
${}^{8}Li$ & ${}^{7}Li$+n & ${}^{13}C$ & ${}^{12}C$+n\\
${}^{9}Li$ & ${}^{8}Li$+n & ${}^{14}C$ & ${}^{13}C$+n\\
${}^{10}Li$ & ${}^{9}Li$+n & ${}^{15}C$ & ${}^{14}C$+n\\
${}^{11}Li$ & ${}^{9}Li$+n+n & ${}^{16}C$ & ${}^{15}C$+n\\
${}^{8}Be$ & ${}^{4}He$+${}^{4}He$ & ${}^{13}N$ & ${}^{12}C$+p\\
${}^{9}Be$ & ${}^{8}Be$+n & ${}^{14}N$ & ${}^{10}B$+${}^{4}He$\\
${}^{10}Be$ & ${}^{9}Be$+n & ${}^{15}N$ & ${}^{11}B$+${}^{4}He$\\
${}^{11}Be$ & ${}^{10}Be$+n & ${}^{16}N$ & ${}^{15}N$+n\\
${}^{12}Be$ & ${}^{11}Be$+n & ${}^{17}N$ & ${}^{16}N$+n\\
${}^{13}Be$ & ${}^{12}Be$+n & ${}^{13}O$ & ${}^{12}N$+p\\
${}^{14}Be$ & ${}^{12}Be$+n+n & ${}^{14}O$ & ${}^{13}N$+p\\
${}^{8}B$ & ${}^{7}Be$+n & ${}^{15}O$ & ${}^{14}N$+p\\
${}^{9}B$ & ${}^{8}Be$+n & ${}^{16}O$ & ${}^{12}C$+${}^{4}He$\\
${}^{10}B$ & ${}^{6}Li$+${}^{4}He$ & ${}^{17}O$ & ${}^{13}C$+${}^{4}He$\\
${}^{11}B$ & ${}^{7}Li$+${}^{4}He$ & ${}^{18}O$ & ${}^{14}C$+${}^{4}He$\\
${}^{12}B$ & ${}^{11}B$+n\\
\hline
\end{tabular}
\caption{The possible NFC structures for light nuclei.}\label{tab:table}
\end{table}

\begin{table}
\centering
\begin{tabular}{cccccccc}
\hline
Nuclei & $R_{f}(fm)$  & $R_{m}(fm)$ & Ref. \\
\hline
${}^{6}Li$ & $2.217$ & $2.32\pm{0.03}$ & \cite{bibitem27} \\
${}^{6}He$ & $2.583$ & 2.45(10) & \cite{bibitem28} \\
${}^{7}Li$ & $2.374$ & $2.33\pm{0.02}$ & \cite{bibitem27} \\
${}^{8}He$ & $3.100$ & $2.53(8)$ &\cite{bibitem28} \\
${}^{8}B$ & $2.635$ & $2.55\pm{0.08}$ &\cite{bibitem29}\\
${}^{8}Li$ & $2.629$ & $2.583\pm{0.023}$ & \cite{bibitem27} \\
${}^{9}C$ & $2.874$ & $2.71(32)$ &\cite{bibitem30}\\
${}^{10}B$ & $2.647$ & $2.56\pm{0.23}$ & \cite{bibitem27} \\
${}^{11}Li$ & $3.432$ & $3.34_{-0.08}^{+0.04}$ &\cite{bibitem31}\\
${}^{11}Be$ & $3.027$ & $3.039\pm{0.038}$ & \cite{bibitem32} \\
${}^{12}B$ & $3.016$ & $2.723\pm{0.050}$ & \cite{bibitem27} \\
${}^{12}C$ & $2.848$ & $2.48\pm{0.08}$ & \cite{bibitem27} \\
${}^{13}B$ & $3.193$ & $2.746\pm{0.050}$ & \cite{bibitem27} \\
${}^{14}Be$ & $3.736$ & $3.36\pm{0.19}$ & \cite{bibitem27} \\
${}^{14}B$ & $3.383$ & $3.00\pm{0.10}$ & \cite{bibitem27} \\
${}^{14}N$ & $3.027$ & $2.61\pm{0.10}$ & \cite{bibitem27} \\
${}^{16}O$ & $3.194$ & $2.63\pm{0.06}$ & \cite{bibitem27} \\
\hline
\end{tabular}
\caption{The comparison of the calculated values of the nuclear fractal mean density radii ( here we use $R_{f}$ to stand for the nuclear fractal mean density radii for convenience.) using
the model of nuclear fractal structure with the experimental results of root-mean-square (rms)
matter ($R_{m}$) radii. The cluster structures of the nuclei listed
in here are same as those in TABLE I.}
\end{table}

In order to describe the essential feature of the nuclear fractal self-similarity symmetry, fractal dimension is one of the basic geometrical parameters. It¡¯s related to the nuclear homogeneity and the ingredients in nucleus, which reflect the nuclear irregular degree. It can be seen that the values of the dimensions are much close to 3, if the mean densities of every NFC in one nucleus and the mean density of this nucleus are all not much different and the irregular degrees of the nuclear structures are low. Based on the NFC structures (TABLE I) and the values of the dimensions (FIG. 2), some interesting results may be discussed. For isobars, because the degree of homogeneity of the stable nuclei is higher than the nuclei far from the stable line generally, the values of the fractal dimension for the former are greater than for the latter. The fractal dimensions of isobars are comparative or same, if the NFC structure types of these nuclei are similar. For example, the fractal dimensions of the isobars with two-body structure type, (${}^{5}Li$, ${}^{5}He$), (${}^{8}Li$, ${}^{8}Be$, ${}^{8}B$), (${}^{10}Be$, ${}^{10}B$), (${}^{11}C$, ${}^{11}B$), (${}^{13}N$, ${}^{13}C$) and (${}^{16}C$, ${}^{16}N$), are almost same respectively. For nuclei with same $A$, the fractal dimensions of the nuclei with three-body structure, such as, ${}^{6}He$, ${}^{8}He$, ${}^{10}C$, ${}^{11}Li$ and ${}^{14}Be$, are distinctly lower than the nuclei with two body structure type. The values of fractal dimension decrease due to the increase of nuclear mass number, which may be associated with the scale $r_{s}$. The fractal scale variables amount to the resolution of the measurement \cite{bibitem33}, which are related to the scale relativity. For instance, if the half of the maximum size of one nucleon serves as the fractal scale, using a 3-dimension bulk whose radius equals to this scale to cover a nucleon, the dimension of one nucleon is 3. However, along with the increase of the nuclear mass number, the nuclear structure becomes more irregular. Therefore, using this 3D structure to measure the nucleus, the dimension of this nucleus is lower than 3, which reflects the information of the nuclear irregular structures. In brief, as the results of the effects of all structure variables and binding energy, the fractal dimension can generally describe the nuclear structure well. Besides, in the process of nuclear synthesis $D_{f}$ is a dynamical parameter, which when $D_{f}$ gets a fixed value, then it explains the existence of relatively stable nuclei.
\begin{figure}
\includegraphics[width=0.45\textwidth]{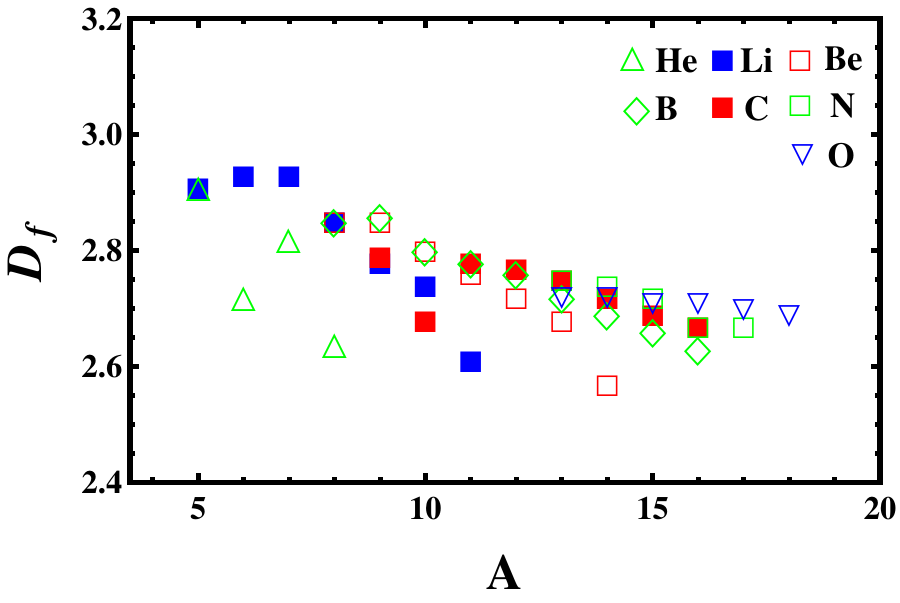}
 \caption{\label{Fig_2} the values of the fractal dimensions of the nuclei.}
\end{figure}

\begin{figure}
\includegraphics[width=0.45\textwidth]{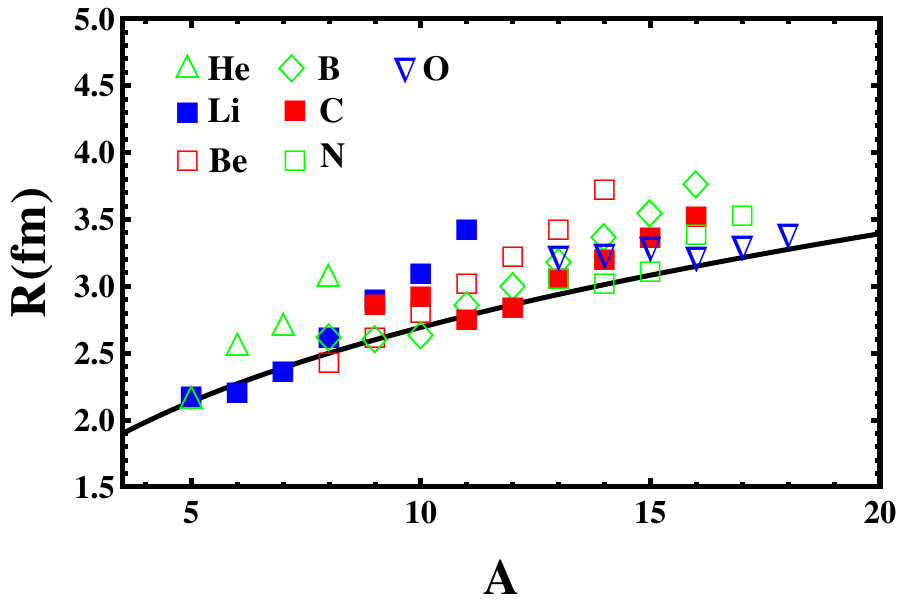}
\caption{\label{Fig_3} the comparison of the calculated values of radii using the model of nuclear fractal structure with the liquid drop model. The color symbols stand for the nuclear fractal mean density radii.}
\end{figure}

The nuclear fractal mean density radii depending on the scale variable $r_{s}$ in the process of measurement can represents the nuclear size statistically, which are mainly determined by the binding energy and NFC structure types and related to the fractal dimension by the expression (5). For isobars, the more is the binding energy, the less is the radii. Generally, as showed in FIG. 3, for stable nuclei and the nuclei very near the stable line, the values of the nuclear fractal mean density radii are approaching the line of $R_{L}$ depending on liquid drop model. When the nuclei are far from the line of stability, the values of the nuclear fractal mean density radii will be greater than $R_{L}$ and this will increase as the deviation from the line of stability increases. In some special cases, the nuclei having same NFC structure and having almost same binding energy, such as ${}^{5}Li$ and ${}^{5}He$, the radii are nearly same. The same situations arises in the pair of isobars (${}^{8}B$, ${}^{8}Li$), (${}^{9}Be$, ${}^{9}B$) and (${}^{13}C$, ${}^{13}N$). As a matter of fact, the liquid drop model cannot calculate the radii of the nuclei far from stable line successfully, as they have much larger radii contrary of the predictions by liquid drop model. On the contrary, the present fractal structure model can successfully predict the radii of such nuclei far from the line of stability. Moreover, the comparison of the calculated values of the nuclear fractal mean density radii using the model of nuclear fractal structure with the experimental results of root-mean-square (rms)
matter ($R_{m}$) radii are listed in TABLE II, where, $R_{m}=\sqrt{\frac{Z}{A}R_{p}^2+\frac{N}{A}R_{n}^2}$ \cite{bibitem31}. And, for the stable nuclei, the results of the calculated radii both using the liquid drop model and the present method are larger than the experimental results of root-mean-square (rms) matter ($R_{m}$).

With regard to the parameter $\alpha$ introduced in the second part, it can be solved as same time as the fractal dimension $D_{f}$ and the nuclear fractal mean density radius $R$ through the three equations explained in the end of the second part, which vary from 0 to 1 in present calculations generally corresponding to the nuclei from stability to unstability.

In summary, the NFC structures are determined by the interactions within nuclear systems and that the fractal dimension can generally describe such structural features. The nuclear fractal mean density radii represent approximately the nuclear size, which are associated with the scale variables. Actually, the relations among the nuclear structure geometric variables are complex and correlative.

\section{Conclusions}
In the present work, we consider the importance that the nuclear irregular structure properties and the self-similarity characteristic may be the intrinsic aspects of nuclear structure properties. For the description of nuclear geometric properties, nuclear fractal dimension is an irreplaceable variable similar to the nuclear radius. Compared with the liquid drop model, it is a feature that the fractal description can reflect the important characteristics of the NFC structures especially for describing the nuclei far from the line of stability and $\alpha$-cluster nuclei. Similar to the liquid drop model, the present model can get the same results of the relation between $A$ and $Z$ on the $\beta$ stable line for the light stable nuclei and cannot predict the existence of the magic nuclei. For heavier nuclei, more realistic formulae for the depth of nuclear potential energy are needed to be obtained whose fractal structure is more complicated. These heavy nuclei may have multilevel fractal structure with several clusters in every level. An anisotropic description for the nuclear fractal structure and the scale dependent properties of the nuclear fractal system would be further studied in future. Further studies will be focused on combination between the properties of the nuclear fractal structure and the quantum mechanics.

\section{Acknowledgments}
This work was partially supported by the Major State Basic Research Development Program of China (973 Program: New physics and technology at the limits of nuclear stability) and the Directed Program of Innovation Project of the Chinese Academy of Sciences with Grant No. KJCX2-YW-N44. The author Wei.Hu.Ma thanks the University of Chinese Academy of Sciences for the education. And one of the authors S.Mukherjee acknowledges the financial help from TWAS-UNESCO, Italy, for visiting Institute of Modern Physics-Chinese Academy of Science, Lanzhou, China for this work.

\end{spacing}
\end{document}